# Ballistic-aggregated Carbon Nanofoam in Target-side of Pulsed Laser Deposition for Energy Storage Applications

Subrata Ghosh[1*], Massimiliano Righi[1], Andrea Macrelli[1], Giorgio Divitini,[2] Davide Orecchia[1], Alessandro Maffini[1], Francesco Goto[3], Gianlorenzo Bussetti[3], David Dellasega[1], Valeria Russo[1], Andrea Li Bassi[1], Carlo S. Casari[1*]

[1] *Micro and Nanostructured Materials Laboratory — NanoLab, Department of Energy, Politecnico di Milano, via Ponzio 34/3, Milano, 20133, Italy*

[2] *Electron Spectroscopy and Nanoscopy, Istituto Italiano di Tecnologia, via Morego 30, Genova, 16163, Italy*

[3] *Solid Liquid Interface Nano-Microscopy and Spectroscopy (SoLINano-Σ) lab, Department of Physics, Politecnico di Milano, Piazza Leonardo da Vinci 32, 20133 Milano, Italy*

**Abstract**: In pulsed laser deposition, along the traditionally exploited deposition on the front-side of the plasma-plume, a coating forms on the surface of the target as well. For reproducibility, this residue is usually cleaned and discarded. Here we instead investigate the target-side coated materials and employ them as a binder-free supercapacitor electrode. The ballistic-aggregated, target-side nanofoam is compact and features a larger fraction of $sp^2$-carbon, higher nitrogen content with higher graphitic-N and lower oxygen content with fewer COOH groups than that of diffusive-aggregated conventional nanofoams. They are highly hydrogenated graphite-like amorphous carbon and superhydrophilic. The resulting symmetric micro-supercapacitor delivers higher volumetric capacitance of 522 mF/cm$^3$ at 100 mV/s and 104% retention after 10000 charge-discharge cycles over conventional nanofoam (215 mF/cm$^3$ and 85% retention) with an areal capacitance of 134 µF/cm$^2$ at 120 Hz and ultrafast frequency response. Utilizing the normally discarded target-side material can therefore enable high performing devices while reducing waste, cost and energy input per usable product, leading towards a greater sustainability of nanomaterials synthesis and deposition techniques.

*Keywords: Pulsed Laser Deposition, nanocarbons, Plasma plume, micro-supercapacitor, Energy storage*

Corresponding author email: subrata.ghosh@polimi.it (S.G.) and carlo.casari@polimi.it (C.S.C.)

**Introduction**

Cultivating nanomaterial thin films with optimized morphology and properties has received tremendous attention from the community for their potential application since the early days of nanotechnology.[1][2][3] The development of various deposition techniques and their advances are direct consequences. Among the widely used techniques, Pulsed Laser Deposition (PLD) has established its ground for preparing materials from carbon to metal oxides, from compact film to 3D porous structures, heterostructures to layered structures and so on.[4][5][6] In the standard geometry, the substrates on which the desired film is to be grown are generally placed in front of the target material, so that the ablated species travel forward with high kinetic energy and get deposited on the substrate (Figure S1a).[7] In certain ranges of process parameters, at high deposition pressure under background gas(es), for some laser fluence (energy per pulse per unit area) values, part of the ablated species are redeposited on the target too (Figure 1b). The species formed inside the plasma plumes are mostly excited atoms, dimers etc, which assemble into materials on the substrate. Indeed, in the case of laser ablation of a graphite target, it has been reported that $sp$-hybridized carbon atomic wires have been observed on the vicinity of the highly oriented pyrolytic graphite (HOPG, $sp^2$-hybridized) target itself.[8] For the reproducibility of the desired conventional materials, the traditional approach is cleaning out the residue or discard materials from the target after every single synthesis run by laser cleaning and/or chemical cleaning. Very few studies consider the film that can be grown on a substrate on the target side, formed by backward ablated species. In order to investigate potential uses for this inevitable by-product of the PLD process, here we investigate: how the backward-ablated species form a structure if a substrate is provided, what the morphology and properties of such film are, how do they differ from conventional materials grown in PLD and whether there is any potential application.

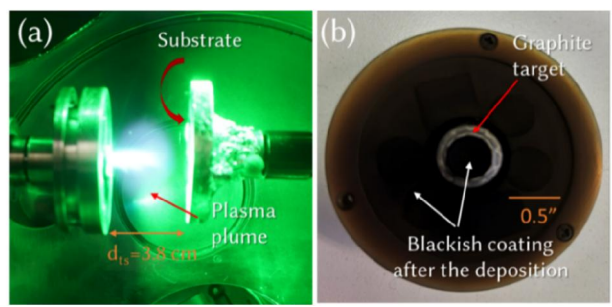

*Figure 1: Optical micrograph of (a) plasma plume during the deposition of nanostructures by pulsed laser deposition with 532 nm laser and (b) target holder after the deposition with the blackish coating. Here, the substrate is placed on the rotating substrate holder in front of the plasma plume and the target is also rotated without any translational motion.*

Applications-wise, carbon-based materials are quite popular as supercapacitors or electrochemical capacitors, where supercapacitors have higher energy density than conventional capacitors and higher power density than conventional batteries. Currently, micro-supercapacitor or thin film-based supercapacitors have received significant attention as portable, wearable, lightweight and miniaturized energy storage devices and *a.c.* line filtering applications.[9][10] However, for carbon-based materials, the major drawbacks are the hydrophobicity in aqueous electrolyte,[11] limited density of states,[12] and ten-to-hundred times lower specific capacitance than that of pseudocapacitive materials[13]. Moreover, to fabricate the electrode, most of the carbon-based materials are mixed with a conducting agent and binder, which limits the conducting pathways for the electrolyte ions, cycle stability and sometimes leads to the electrode materials becoming resistive. This fact also limits the ability to operate at relatively low working frequencies (below 1Hz). Thus, the design of binder-free, conducting agent-free, and self-standing suitable electrode materials for high-frequency micro-supercapacitor applications is necessary. With this motivation, we placed the graphite target on a target holder with large diameter, and the empty spaces in the holder crown were filled with substrates (Figure S1a-b). Carbon nanofoam at target-side (CF-T) was synthesized at room temperature by PLD and conventional one (CF-C) is deposited in same growth run (Figure S1c-d). Thorough structural and morphological investigations and utilization testing as supercapacitor electrode were carried out to the best of our knowledge the first time, showing excellent cycle stability and ultrafast frequency response.

**Results and Discussion**

Carbon nanofoams on target-side and substrate-side are synthesized using pulsed laser deposition at room temperature. The synthesis procedure is detailed in the experimental section. Figure 2 displays the field-emission scanning electron micrographs of the target-side nanofoam (CF-T) grown on Si. A similar morphology is observed for the sample grown on carbon paper (Figure S2a) and used for supercapacitor testing. The tree-like CF-T have an average thickness of 8.6 µm (Figure 2b-c) and are relatively densely packed, with a porosity, i.e. a volumetric void fraction, of 79% . The mass density of CF-T is estimated to be 0.45 g/cm$^3$ (EDDIE software,[14] fitted spectra are provided in Figure S2d). Moreover, the binder-free nanofoam surfaces are found to be superhydrophilic, which facilitates efficient electrolyte ion adsorption and desorption, effective ion transport, and ion transfer kinetics upon its utilization as active energy storage electrode (inset of Figure 2b). The electron micrographs (Figure 2c and 2g) demonstrate a good conformal growth capability, necessary for substrates with complex geometry. The summary of morphology and thickness of the sample is presented in Figure S2(b). From EDX (Figure 2d) and XPS (Figure 3a), the chemical composition is found to be predominantly carbon, bulk- and surface-functionalized nitrogen (as $N_2$-$H_2$ was used as background gas in deposition), and surface-functionalized oxygen. In contrast, CF-C, with an average thickness (20 µm) is found to be taller compared to the CF-T, more porous (volumetric void fraction ~ 93%) and lighter (mass density ~ 0.14 g/cm$^3$). No signature from nitrogen is found for the CF-C from EDX measurement. Figure 3(a) and 3(b) show TEM images for CF-T and CF-C, respectively. The samples are qualitatively similar, without any ordered structures at lattice scale. Particles in a 5-15 nm size range assemble into highly porous micron-scale aggregates, clearly visible in the high-angle annular dark-field scanning transmission electron microscopy (HAADF-STEM) images (Figure 3(c) and 3(d)). The electron energy loss spectra (EELS) of CF-T and CF-C (Figure 3(c)) display a carbon edge that is similar to the amorphous lacey carbon support of the TEM grid, confirming the nature of the film. Furthermore, EELS supports the presence of nitrogen in CF-T, identifiable from the edge at ~400 eV. The edge is much less prominent and close to noise level in CF-C, in accordance with the EDX elemental analysis.

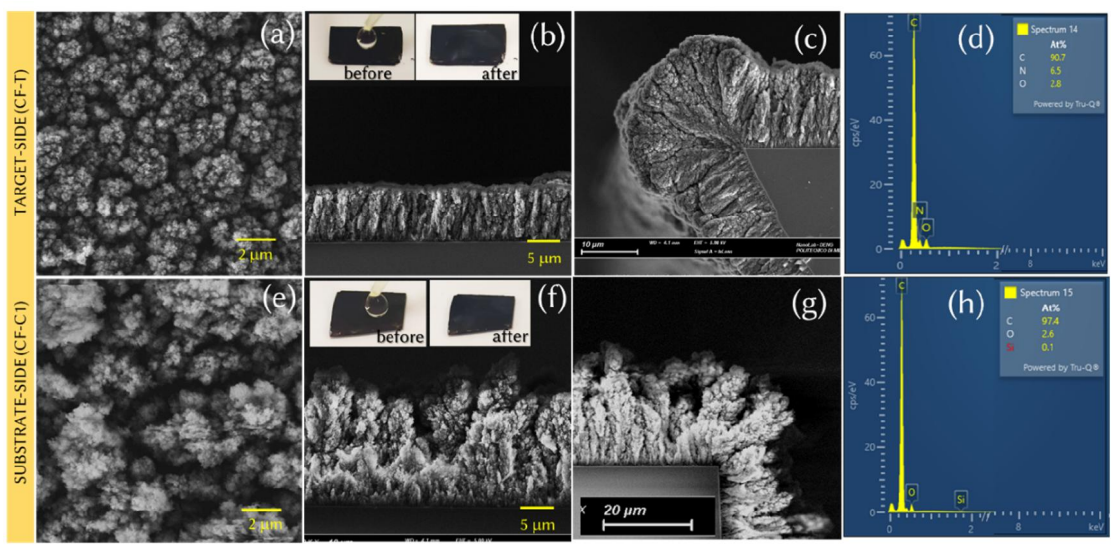

*Figure 2:* (a) Top-view and (b-c) cross-sectional view and (d) EDX spectra with elemental quantification of CF-T. (e) Top-view and (f-g) cross-sectional view and (h) EDX spectra with elemental quantification of CF-C. The inset of Figures 2(b) and 2(f) are the photographic images of the water droplets on CF-T and CF-C, respectively, where volume of water droplet was 10 µl.

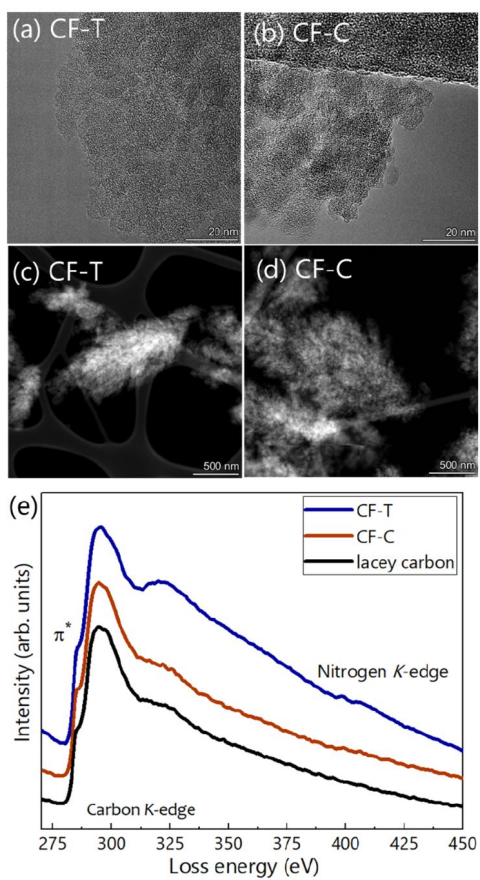

*Figure 3:* Transmission electron micrograph of (a) CF-T and (b) CF-C. STEM-HAADF (high-angle annular dark field) images of (c) CF-T and (d) CF-C. (e) Electron energy loss spectra of CF-T and CF-C along with lacey carbon as a reference.

The XPS survey spectra of CF-T and CF-C are shown in Figure 4a, and the atomic concentration of carbon, nitrogen and oxygen is provided in the table in inset of Figure 4a. The presence of oxygen and nitrogen - functional groups is not only known to improve the wettability of the carbon materials but also to contribute to the quantum capacitance and pseudocapacitance when employing the material as an energy storage electrode.[12] The elemental analysis confirmed the presence of 10 at.% of nitrogen in CF-T, and a lower 6.4 at.% in CF-C. The higher amount of nitrogen content in CF-T is due to the higher reactivity of nitrogen species inside the plasma plume (more radicals etc.). In the plasma plume, there is also a possibility of nitrogen ions forming and depositing on the target-side, where the path is shorter and hence the average number of collisions with the background gas species is lower. They could be instead prevented to reach the substrate-side, recombining into $N_2$ without significant film incorporation. The fact that on the substrate-side nitrogen can be seen with surface-sensitive XPS and not with EDX suggests that substrate-side nitrogen could be predominantly surface contaminants. On the other hand, CF-C contains relatively higher atomic oxygen concentration (30.9%) compared to CF-T (26.4%) due to the higher volumetric void fraction of CF-C.

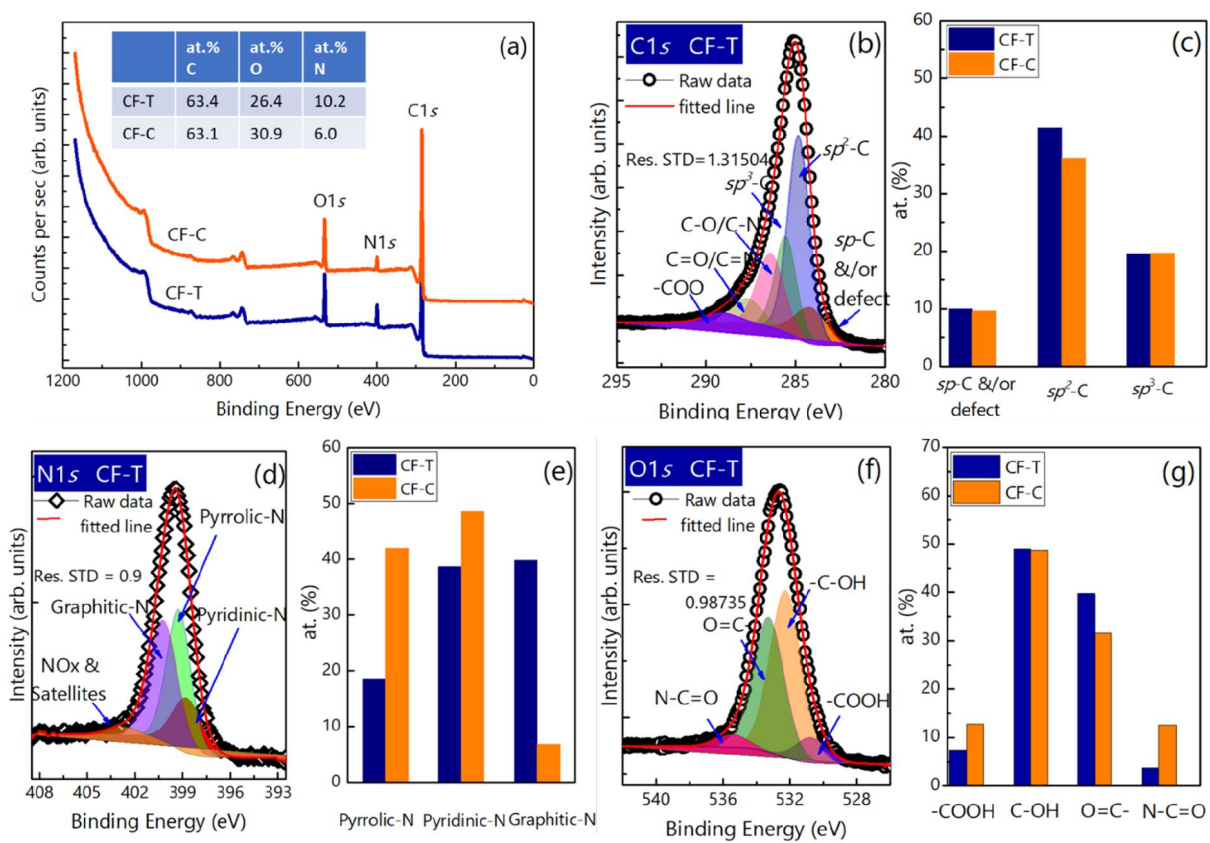

*Figure 4:* (a) Survey scans with at. % of C, O and N *of CF-T and CF-C at inset. High resolution (c) C1s* spectra with (d) relative contribution of carbon components. (e) High-resolution N1s with (f) relative at. % concentration of individual functional groups, and (g) High-resolution O1*s* spectra with (h) relative at. % concentration of individual functional groups of CF-T and CF-C.

The full width at half maximum (FWHM) of the C1$s$ peak of both CF-T and CF-C is found to be around 2.2 eV (Figure 4b), which is much higher than that of highly oriented pyrolytic graphite (HOPG) (0.8 eV). The broadened C1$s$ peak indicates the CF is amorphous in nature, which is investigated further by Raman spectroscopy and discussed later, and in agreement with the TEM and EELS measurements. The best fit for high-resolution C1$s$ for our carbon nanofoams is obtained by introducing the additional peak observed at around 283.8 eV along with $sp^2$-C (~284.8 eV), $sp^3$-C (~285.5 eV) and other functionalized carbon bonds (Figure 4b). The high-resolution C1$s$ spectra can be fitted by two carbon peaks ($sp^2$-C and $sp^3$-C) and functionalized carbon peaks. However, the fitting with those deconvoluted peaks leads to higher residual standard deviation at lower energy (see Figure S3). For the carbon nanostructures formed by plasma-assisted growth, the peak at lower binding energy (~ 283.8 eV) is assigned to vacancy defects.[15] The peak at lower binding energy has been reported for the defected HOPG, prepared by sputtering with 60 eV Ar$^+$ ions. Supported by density-functional theory calculations, a chemical shift towards a lower binding energy for the all carbon atoms surrounding a single vacancy in hexagon matrix has been reported with respect to the C1$s$ peak of a perfect HOPG (284.5 eV).[16] In the case of amorphous carbon nanofoams grown at more than 30 Pa under Ar gas, onion-like-carbon is found to be embedded in the carbon nanofoams matrix from transmission electron microscopy.[17] The presence of onion-like-carbon indicates the existence of heptagon-pentagon structure in the carbon nanofoams matrix.[17] This peak is also assigned as defected graphitic peak for the amorphous carbon thin film prepared by ion-beam sputtering technique.[18] Moreover, the peak at around 283.8 (±0.3) eV obtained after deconvoluting C1$s$ peak of carbyne-like carbon films[19], nanothick amorphous carbon[20], and nanostructured carbon materials is assigned to the $sp$-C.[21] In our case, as carbon nanofoams is prepared at room temperature by laser ablating the graphite in PLD, we observed $sp$-carbon in agreement with our previous work.[22] The fitting of the $sp$-C along with the $sp^2$-C and $sp^3$-C is also reported for the $sp$-bonded carbon chains on the graphite surface prepared by femtosecond PLD[23] and the amorphous carbon[21]. It has been reported[21] that $sp$-C content in amorphous carbon decreases when it is annealed. Therefore, the peak at 283.8 eV is attributed to the $sp$-C and/or vacancy defect present in our carbon nanofoams. To validate the $sp$-C presence in the structure, the Raman spectra were collected for longer accumulations, as discussed later. Indeed, more investigation on $sp$-C binding energy and its contribution to total C1$s$ will be the subject of further experimental and theoretical investigations, but not the focus of this current research. Among the samples, CF-T shows relatively higher $sp^2$-C and lower $sp^3$-C content than that of CF-Cs (Figure 4c), which could have impact on the electrochemical performances.

The high-resolution N1$s$ spectra of CF-T (Figure 4d) is deconvoluted into four peaks: pyridinic-N, pyrrolic-N, graphitic-N and NO$_x$ or satellite peaks. From the analysis, the graphitic N-content at 400.2 eV in CF-T sample is found to be higher than that of CF-Cs (Figure 4e). On the other hand, CF-C has the highest content of pyridinic-N at 399.2 eV and pyrrolic-N at 399.9 eV. It can be also seen from the high-resolution O1$s$ spectra (Figure 4f) that the CF-T and CF-C have different content of oxygen functional groups on the surface as they have grown at different distances and zone in the plasma plume. From Figure 4g, it is clear that the CF-C contains a higher amount of -COOH groups compared to the CF-T. The -COOH group is mostly unstable and leads to poor cycle stability of electrodes for longer charge-discharge cycles.[24] The high-resolution spectra of N1$s$ and O1$s$ for CF-C with deconvoluted peaks are provided in Figure S4.

The Raman spectra of CF-T and CF-C (Figure 5a) consist of a large band related to the $sp^2$ signal, comprising two components, namely D peak at ~1368 cm$^{-1}$ and G-peak at ~1576 cm$^{-1}$, along with very a weak $sp$-band (inset of Figure 5a) and a small, modulated bump in the region of 2400-3200 cm$^{-1}$. A similar feature in the Raman spectra samples grown on carbon paper is noticed (Figure S5). The weak *sp*-C feature obtained in Raman spectra[22] is in agreement with the very low *sp* content extracted from XPS data. Figure 5a also indicates the presence of photoluminescence background, which is usually related to the presence of hydrogen in the sample.[25] It is noteworthy to mention that CF-T has higher photoluminescence than CF-C. To estimate the hydrogen content in the sample from these spectra, we tentatively use the formula reported in Ref. [25], which investigated a broad range of hydrogenated amorphous carbon (*a*-C:H) material. The quantitative formula used is $H[at.\%] = 21.7 + 16.6 \log\left[\frac{m}{I_G}(\mu m)\right]$[25], where m is the slope of first order Raman spectra within the region of 900 to 1900 cm$^{-1}$ (Figure 5b) and $I_G$ is the intensity of the G peak (extracted G-peak height from Raman spectra, Figure 5b). The photoluminescence background $\left(\frac{m}{I_G}\right)$ for CF-T and CF-C was found to be 4.32 and 3.58 μm, respectively, close to the typical value for highly-hydrogenated graphite-like amorphous carbon (GLCHH) which is reported to be around 5 μm.[25] The higher the slope and/or ratio of $\frac{m}{I_G}$, the sample is more hydrogenated. This fact suggests that CF-T is more hydrogenated than CF-C and this would agree with the fact that on the target side most reactive species are available. The estimated hydrogen content of CF-T and CF-C, using the above-mentioned formula, are 32.2% and 30.9%, respectively. On the other hand, XPS analysis reports that both CF-T and CF-C contain around 19% $sp^3$-C (C-C $sp^3$ + C-H $sp^3$). Moreover, to infer the type of hydrogenated carbon from Raman spectra one needs to also take into account the G-peak position and the $I_D/I_G$ intensity ratio as a function of the estimated hydrogen content, as reported in Ref.[25] and adapted in Figure 5c. Using our data, we determine that our carbon nanofoams do not fall into the extrapolated line of *a*-C:H (red line in Figure 5c) of Ref. [25]. This is an indication that the presence of nitrogen in carbon nanofoams, while not giving any contribution to photoluminescence, influences the Raman signal, namely the D and G peaks. Actually, it is known that nitrogen has a significant role, in clustering and bonding environment of different hybridized carbon.[26] We tentatively use the three-stage model for amorphous carbon nitride containing hydrogen reported in Ref. [26] and adapted in Figure 5d. Plotting the G-peak position and $I_D/I_G$ of our samples with respect to the $sp^3$-C content (obtained from XPS: 19.51% for CF-T and 19.66% for CF-C), it appears that our carbon nanofoams falls in the non-uniqueness region of stage-II of the model. $N_2$-$H_2$ background gas is used for the deposition and the presence of nitrogen content in the carbon nanofoams (~10 at% in CF-T and 6% at% in CF-C) is confirmed from XPS.

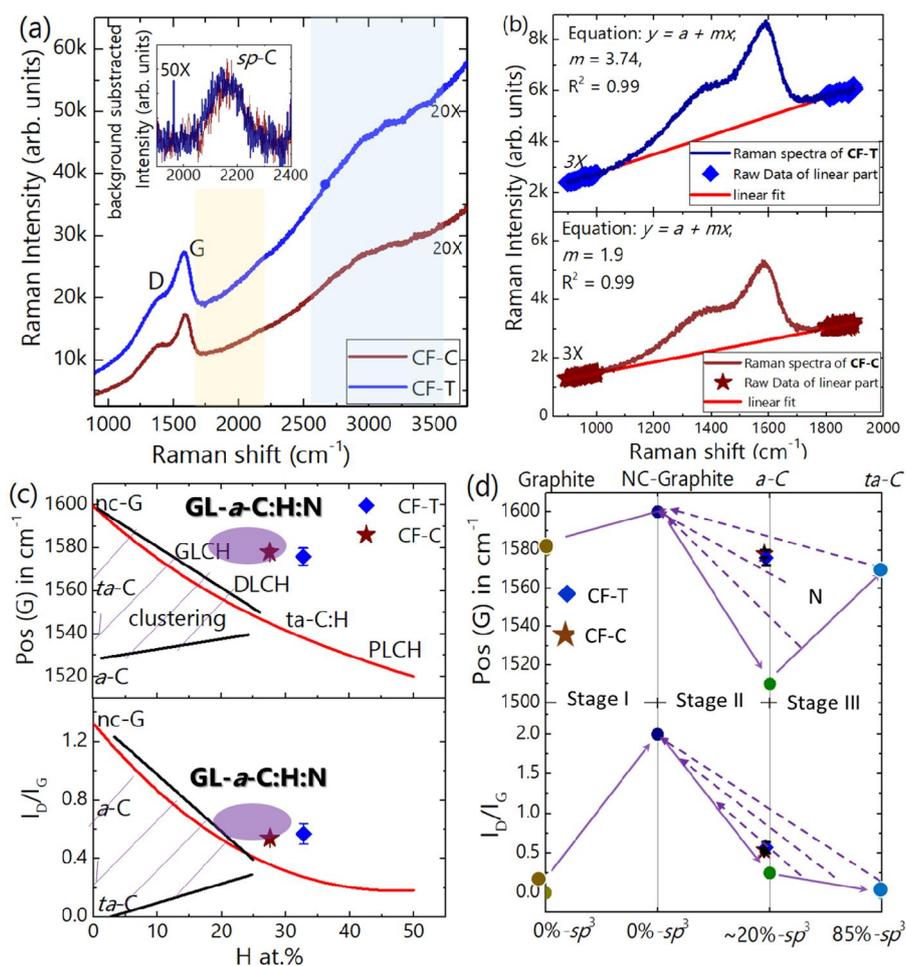

*Figure 5:* (Color online) Raman spectra of CF-T with the laser used of 514 nm with photoluminescence background. (b) First-order Raman spectra of CF-T at top and CF-C at bottom with the linear fit of Raman data of linear regions in the range of 900 to 1900 cm$^{-1}$ (◆ and ★ in the graph belongs to CF-T and CF-C, respectively). Plot of G-peak position and $I_D/I_G$ of carbon nanofoam (c) with respect to the hydrogen-content and (d) in Ferrari-Robertson's amorphization trajectory model. The red and black solid line in (c) is the fitted data extracted from the extrapolated fitted line from the Ref.[25] using the WebPlotDigitizer software authored by Ankit Rohatgi. Shaded regions are non-uniqueness region at low hydrogen-content. GLCH, DLCH, ta-C:H and PLCH represent graphite-like, diamond-like, tetrahedral and polymeric hydrogenated amorphous carbon (*a*-C), respectively. *nc*-G, *ta*-C and GL-*a*-CHHN represent nanocrystalline graphite, tetrahedral *a*-C and highly hydrogenated graphite-like *a*-C nitride. In Figure 5(d), solid lines guide the eye to indicate the transition of one stage to another stage and the dashed arrow lines represent the non-uniqueness region due to the incorporation of nitrogen in amorphous carbon.[26]

The higher amount of heteroatoms like hydrogen, oxygen and nitrogen for CF-T is also reflected in the Raman spectra as we see broadened FWHM of D-peak (281.7±34.5 cm$^{-1}$) and G- peak of (127.6±6.6 cm$^{-1}$) compared to that of CF-C (249.5±17.8 cm$^{-1}$ for D-peak and 121.4±1.5 cm$^{-1}$ for G-peak). Thus, according to the classification of hydrogenated *a*-C and *a*-C:N,[25],[26] and the above analysis, the carbon nanofoams prepared by PLD can be defined nitrogen-containing graphite-like *a*-C:H (GL-*a*-C:H:N). A systematic investigation on the GL-*a*-C:H:N can be the subject of further research.

The distinctively columnar morphology of CF-T films is indicative of a ballistic aggregation, in contrast with the diffusive aggregation typical of fractal-like ultra-low density nanofoams deposited at higher background pressure and lower fluence.[23],[27] Since the ablated material is ejected with a non-zero momentum toward the substrate, one can argue that the CF-T deposits are grown by species that have undergone a back-reflection caused by single or multiple large angle collisions with background gas atoms and molecules, likely occurring in the vicinity of the ablation region. For this reason, the mean path traveled by the species that arrive on the target side can be considerably shorter than that of species landing on the substrate. In a ballistic regime, the energy lost by the ablated species is proportional to the number of collisions with the background gas: the longer the traveled path, the lower is the kinetic energy of the species that contribute to the film growth. As a result, the morphology of the CF-T deposits is indeed less porous, containing higher hydrogen and nitrogen-content than CF-C ones, hinting to a lower kinetic energy for the conventional deposition compared to the target-side. It is worth noting that it is not straightforward to obtain the same morphology of CF-T in a conventional configuration by simply reducing the target-to-substrate distance. Indeed, such a configuration would imply that the film is grown by a mixture of very energetic species that have undergone few or no collisions and low energy species resulting from many small angle collisions. This difference certainly is expected to have an impact on their performance as an active material for desired applications.

Figure 6(a) shows the cyclic voltammogram (CV) of CF-T symmetric micro-supercapacitor device at the scan rate ranging from 20 mV/s to 1000 mV/s. The shape of CV is quasi-rectangular with a mirror-image feature and maintained its shape with respect to the scan rate indicating excellent capacitive behavior of the device within 0 to 0.7 V. The deviation from a perfect rectangular CV shape is due to the pseudocapacitive contribution from the oxygen and nitrogen functional groups present on the surface of nanofoam. The areal/volumetric capacitance of the microdevice is estimated and plotted with respect to the scan rate in Figure 6(b). The CF-T microdevice is capable of providing higher volumetric capacitance (522 mF/cm$^3$) with higher rate performance of 64% at 1000 mV/s compared to 100 mV/s than that of the CF-C (215 mF/cm$^3$, 54% retention). A comparative volumetric CV profile of CF-T and CF-C device at the scan rate of 100 mV/s and current dependent charge-discharge profile of CF-T devices is shown in Figure 6(c) and 6(d), respectively. From charge-discharge, the estimated volumetric (areal) capacitance of CF-T device obtained at 175 μA current is 463 mF/cm$^3$ (796 μF/cm$^2$) and it retains 74% of its capacitance at higher current of 500 μA. Whereas the CF-C device delivers the volumetric (areal) capacitance of 162 mF/cm$^3$ (659 μF/cm$^2$) at 175 μA current and retains 68% of its initial capacitance at 500 μA current. The CV of CF-C at different scan rates, plot of normalized capacitance with respect to the scan rate and charge-discharge profile are provided in Figure S7a-c of supporting information. It is important to note that CF-T microdevices deliver better cyclic stability of 104% up to 10000 charge-discharge cycles (Figure 6e) and maintain a similar CV profile, in contrast to the CF-C microdevices (cycle stability ~ 85%). The outperformance of CF-T microdevice compared to the CF-C microdevice in charge-storage can be attributed to the higher $sp^2$-C, lower oxygen content with lower unstable COOH group[24] and higher nitrogen-content with higher graphitic-N.[28]

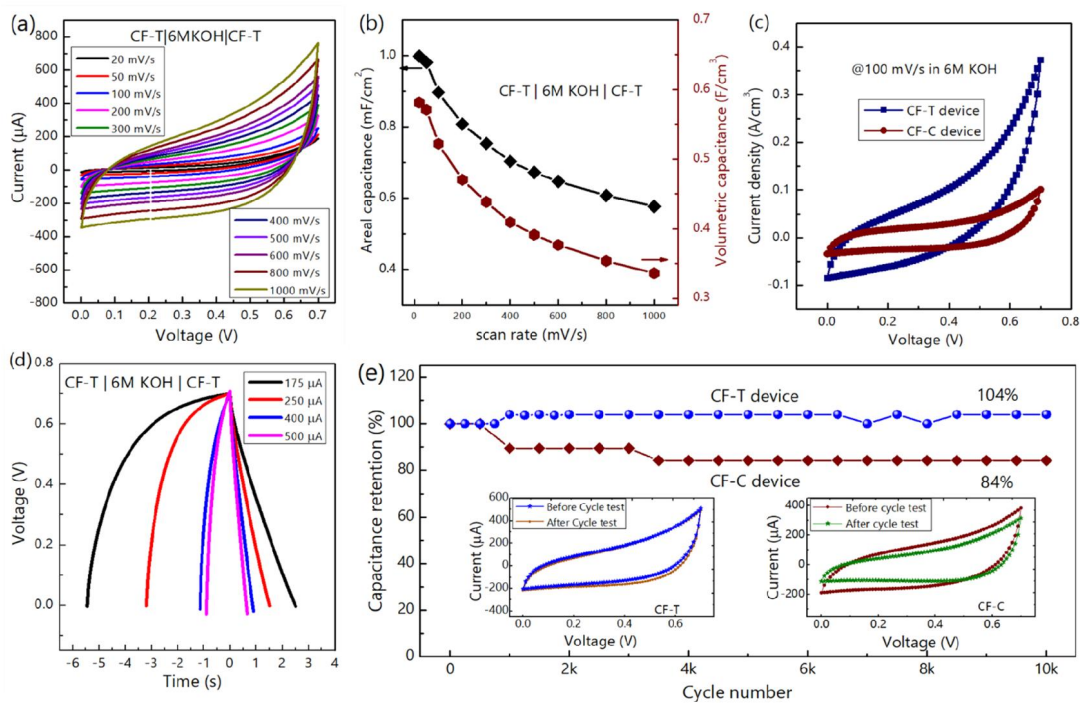

*Figure 6 (color online)*: Electrochemical performances. (a) Cyclic voltammogram and (b) specific capacitance at different scan rates ranges from 20-1000 mV/s for the CF-T symmetric device. (c) comparative cyclic voltammogram and specific capacitance at the 100 mV/s scan rate of CF-T and CF-C symmetric device. Charge-discharge profile of (d) CF-T device at different currents ranges from 175 to 500 µA. Negative sign in time axis represents charging. (e) Cycle stability of both devices at 150 µA with the cyclic voltammogram profile at 500 mV/s of each device before and after 10000 charge-discharge cycles.

The increased capacitance retention of 104% after 10000 charge-discharge cycles for the CF-T device (Figure 6e) can be attributed to the chemical activation associated changes in the electrode materials during the charge-discharge. Such enhanced capacitance retention has already been observed for carbon nanostructures e.g. N-doped mesoporous activated carbon[29], non-stacked reduced graphene oxide.[30] For the N-doped mesoporous activated carbon,[29] 109% capacitance retention after 5000 cycles was attributed to the gradual activation and enhanced wettability over prolonged charge-discharge cycles. Around 103.5% capacitance retention after 15000 charge-discharge cycles for non-stacked reduced graphene oxide is attributed to the increased effective interfacial area between the homogeneous pores of the electrode material and the electrolyte over prolonged charge-discharge cycles. Moreover, structural disorder in the nanoporous carbon is demonstrated to play a key role in capacitance enhancement.[31] To probe insights on the structural changes, we recorded the Raman spectra of the CF-T electrode after the electrochemistry and compared it with the Raman spectra recorded before the electrochemical test (Figure 7a). A significant structural change in terms of position of G-peak and D-peak, FWHM of D-peak and G-peak, and $I_D/I_G$ is clearly observed, which is shown in Figure 7b. It is important to note that the structural changes for CF-C electrode are smaller compared to the changes in CF-T. For example, while FWMH of G(D)-peak for CF-T is broadened by 13 (27) cm$^{-1}$, the change for CF-C is within the unity (2.8 cm$^{-1}$) after the electrochemical test. Thus, from our study, the higher capacitance retention of CF-T device more than 100% over prolonged charge-discharge cycles is attributed to the enhanced structural disorder in CF-T along with the possible chemical activation and wettability improvement.

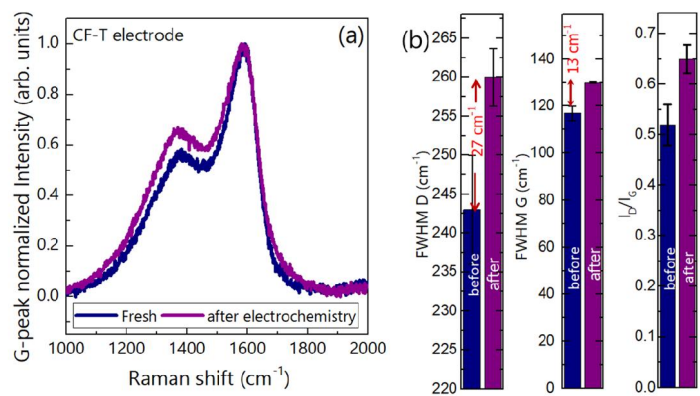

*Figure 7 (color online)*: (a) Raman spectra and (b) Changes in FWHM of D-peak and G-peak, and ID/IG of CF-T before and after the electrochemical test. The change in those parameters is highlighted by red arrow in the figure with quantification.

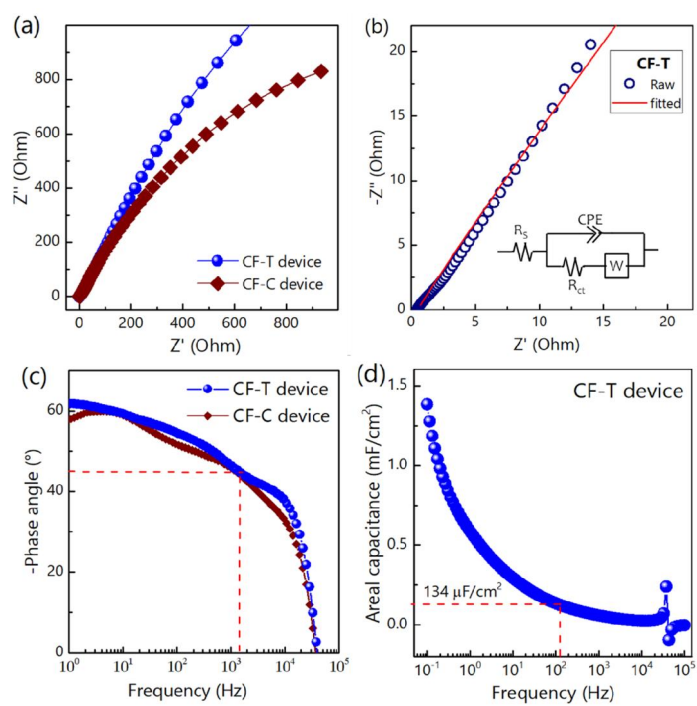

*Figure 8 (color online)*: (a) Nyquist plot of CF-T and CF-C. (b) Nyquist plot of CF-T in high frequency region with the electrical equivalent circuit, shown in the inset, fitted data. (h) Bode plot of CF-T and CF-C device. (i) Frequency dependent areal capacitance of CF-T at 10 mV *a.c.* perturbation

Figure 8a shows the Nyquist plot, where the CF-T profile is steeper than the CF-C, indicating relatively faster ion diffusion and better capacitive behavior. However, the Nyquist plot in the high-frequency zone for both CF-T (Figure 8b) and CF-C (Figure S7d) is very similar. The Nyquist plot is fitted with electrical equivalent circuit modelling (inset of Figure 8b), and the obtained equivalent series resistance and charge-transfer resistance of CF-T from the fitted circuit are 0.5 Ohm/cm$^2$ and 0.4 Ohm/cm$^2$, respectively. At -45°, the corresponding frequency is found to be 1321.9 Hz for the CF-T, which is higher than that of CF-C (1149.75 Hz). The phase angle of CF-T and CF-C is estimated to be -54° and -58.4° at 120 Hz (Figure 8c). Although the phase angle of the carbon nanofoams at 120 Hz is lower than the ideal capacitor (-90°), it is comparable and/or better than many existing reports (Table 1). The frequency-dependent areal capacitance of CF-T device is plotted in Figure 8d. At 120 Hz, areal capacitance is calculated to be 134 µF/cm$^2$ and corresponding areal energy density is 32.8 µFV$^2$/cm$^2$. The relaxation time constant response time for CF-T and CF-C is found to be 0.86 and 0.76 ms, respectively. The resistor-capacitor time constant is found to be 0.12 ms for CF-T. These results indicate that the carbon nanofoam maintain fast frequency response.

*Table 1*: Comparison of micro-supercapacitor devices with frequency response (* represents the calculated value using the data available in the cited reference)

| Electrode materials | Electrolyte | Areal capacitance (CV or CD) in mF/cm$^2$ | Areal capacitance at 120 Hz (EIS) (µF/cm$^2$) | Areal energy density (µFV$^2$/cm$^2$) | Phase angle at 120 Hz (°) | Frequency (Hz) / Relaxation time constant (ms) at -45° | Resistor-capacitor time constant (ms) | Cyclic stability |
|---|---|---|---|---|---|---|---|---|
| CF-T, thickness ~8.6 µm | 6M KOH | 0.9 at 100 mV/s | 134 | 32.8 | -54.04 | 1321.9/0.86 | 0.12 | 104% after 10$^5$ cycles at 125 µA |
| CF-C, thickness ~20 µm | | 0.87 at 100 mV/s | 127 | 31.1 | -58.39 | 1149.75/0.76 | 0.14 | 84% after 10$^5$ cycles 125 µA |
| O-PGF[32] | 0.1 M LiClO$_4$ | 3.8 at 25 µA/cm$^{-2}$ | 755 | 377.5* | -53 | - / 1.46 | 0.98 | 96.7% after 500 µA/cm$^{-2}$ |
| EOG/CNF, thickness ~1 µm[33] | 6M KOH | - | 370 | 149.85* | -81.5 | 22000 / 0.07 | - | 98% after 10$^6$ cycles at 40 mA/cm$^{-2}$ |
| | 1 M TEABF4/AN | - | 160 | - | -80 | 8500 / 0.12 | - | - |
| LCTH[34] | EMIMBF$_4$ ionic liquid | - | 118 | 721* | -61.5 | 1121.3 / 0.89 | 0.7 | 101% at 14000 cycles at 10 V/s |
| VOG[35] | 25% KOH | - | 87 | 43.75* | -82 | 15000 / 0.067 | 0.2 | - |
| CNT[36] | 0.5 H$_2$SO$_4$ | - | 601 | 192* | -81 | - / 0.702 | 0.199 | - |
| CNT[37] | EMImNTf$_2$ | - | 128 | 576* | -80.1 | - / 0.794 | - | - |
| EPDC[38] | 1 m TEABF$_4$ | - | 225 | 703.25 | -76.5 | - / 0.8 | 0.46 | 97.5% after 5000 cycles at 25 µA |
| pristine carbon[39] | 1 M TEABF$_4$ in AN | - | 444 | - | -80 | 890/1.2 | - | - |
| N-doped carbon[39] | | - | 545 | - | -67 | 280/0.56 | - | - |
| P/N-doped carbon[39] | | - | 30 | - | -82 | 13200 / 0.077 | - | - |
| B/N-doped carbon[39] | | - | 99 | - | -83 | 5200 / 0.2 | 0.13 | - |

(CV: cyclic voltammogram, CD: charge-discharge, O-PGF: 3D ordered porous graphene, EOG: edge-oriented graphene, CNF: carbon nanofiber, LCTH: laser-processed carbon-Titanium carbide heterostructure, VOG: vertically oriented graphene, CNT: carbon nanotube, EPDC: Electrospun Polymer-Derived Carbyne, TEABF$_4$: Tetraethylammonium tetrafluoroborate, AN: acetonitrile, EMIMBF$_4$: -Ethyl-3-methylimidazolium tetrafluoroborate, EMImNTf$_2$: 1-Ethyl-3-methylimidazolium bis(trifluoromethane sulfonyl)imide)

Carbon nanofoam was also synthesized using a hydrothermal method[40], a sol-gel method[41], pyrolysis[42], laser processing of graphene oxide[43], chemical vapor deposition,[44] etc. by other groups and used for different applications. All the carbon nanofoams prepared by the above-mentioned method either use high temperature and/or involve multiple steps to obtain the final structure. The signature of *sp*-C from those nanostructures is generally weak, and they are not hydrophilic and lightweight as the carbon nanofoam prepared in our room temperature PLD process. For example, carbon nanofoam synthesized at high temperature by chemical vapor deposition is hydrophobic in nature.[44] Moreover, the carbon nanofoam deposited in frontside and backside of the target ensures the possibility of growing carbon nanofoam with different characteristics in a single production run at room temperature. The room temperature growth enables to grow a film on any flexible substrate. To prove the versatility of our approach, we demonstrate porous nanostructures (on the substrate placed at the front-side of plasma plume), compact (on the substrate placed within the target of target-side), and intermediate nanostructure (on the substrate placed outside the target in target-side) by ablating material from two targets simultaneously by illumination with a single laser, in one synthesis process (Figure S8).

**Conclusion**

Carbon nanofoam is successfully grown both in conventional way (CF-C) and on the target-side (CF-T) in a single production run using pulsed laser deposition at room temperature. The morphology and structural properties of CF-T are found to be different from CF-C in terms of porosity, mass density, thickness, and degree of graphitization, which is attributed to the different aggregation dynamics under complex plasma species' behavior in the forward (diffusive aggregation) and backward (ballistic aggregation) directions. Carbon nanofoams are found to be highly hydrogenated nitrogen-containing graphite-like amorphous carbon and superhydrophilic. Micro-supercapacitors based on CF-T (CF-C) delivered a high areal capacitance of 0.9 (0.87) mF/cm$^2$ and volumetric capacitance of 521 (215) mF/cm$^3$, had 104% (84%) capacitance retention after 10000 charge-discharge cycles, and showed excellent frequency response, with the capability of fast charge delivery. The better performance of uniquely obtained CF-T compared to conventional CF-C is attributed to the compactness, and higher *sp*$^2$-carbon content, higher content of N-functionalities over O-functionalities, higher graphitic-N, and lower COOH. It can be concluded that the proposed strategy of obtaining more than one nanostructure from a single deposition without additional resources, energy, and time, is versatile, of significant value for large-scale production, sustainable, and drives the technology towards the goal of net zero waste. The binder-free, self-supporting active materials can be used for flexible optoelectronics, micro energy storage devices and other applications.

EXPERIMENTAL SECTION

**Synthesis of carbon nanofoam**. Carbon nanofoam was deposited on the target side (CF-T) by placing the substrates, Si and carbon paper, on the 2-inch target holder around the 0.5-inch graphite target. For the conventional deposition, bare carbon nanofoam (CF-C) was deposited on substrates in front of the target at a defined distance of 3.8 cm from graphite. The ablation was performed using a Nd:YAG pulsed ns-laser (2$^{nd}$ harmonic at 532 nm, pulse duration 5-7 ns, repetition rate 10 Hz) for 15 min with a pulse energy of 420 mJ and fluence of 6.5 J/cm$^2$. Prior to the deposition, the chamber was evacuated down to 10$^{-3}$ Pa. Then, the pressure of the deposition unit was maintained at 300 Pa under N$_2$-H$_2$ (95%-5%) environment. Both the target and substrate holder were rotated to uniformly ablate the species and deposit on the substrate, respectively. Optical micrographs of substrate placed on the target side and substrate sides before and after deposition are given in Figure S1.

**Microscopy and Spectroscopy.** The morphology of obtained materials was acquired using a field-emission scanning electron microscope (FESEM, ZEISS SUPRA 40, Jena, Germany) and inLens detector operated in a high vacuum. Electron Dispersive X-ray (EDX) spectra were recorded at the acceleration voltage of 20 kV using a Peltier-cooled silicon drift detector (Oxford Instruments), using Aztec software to acquire data and evaluate the local composition. To calculate the mass density of the film, a MATLAB code called EDDIE software was used.[14] With the input of average height information of the material and the details of substrate reference (Si for our case), one can estimate the mass density of the film using the data obtained from EDX analysis.

Considering the density of graphite ($\rho_{graphite}$) of 2.2 g/cm$^3$ and using the mass density estimated from EDDIE software, the porosity or volumetric void fraction of the grown material is calculated using the formula of $\% \, porosity = \left(1 - \frac{\rho_{ACF}}{\rho_{graphite}}\right) \times 100$.

Transmission electron microscopy was carried out on a ThermoFisher Spectra300 S/TEM operated at 300 kV. TEM images were acquired on a Ceta camera. STEM-HAADF images were collected using a Panther detector, and EEL spectra using a Continuum GIF with a current of ~300 pA and an energy resolution of ~1 eV.

The XPS data were acquired using a non-monochromated x-ray source using a Mg anode (photon energy 1253.6 eV), maintained at a power of 200W. The kinetic energy of the photoemitted electrons was measured using a hemispherical analyzer with a 150mm mean radius, PHOIBOS150 from SPECS GmbH. The spectra were acquired with a pass energy of 20 eV, with an energy resolution of 0.9 eV (FWHM). The pressure in the measurement chamber during the experiments is about 1x10$^{-10}$ Torr. Peaks were fitted after Shirley background subtraction using CasaXPS software, and at.% of elemental compositions were extracted from peak area ratios after correction by Scofield relative sensitivity factors (C = 1.0, N = 1.77, O = 2.85).[45] For C1s, the asymmetric *sp*$^2$-C peak (calibrated at 284.8 eV) is fitted with Gaussian-Lorentzian lineshape (GL(30)) with asymmetric factor (T200)[46] and other symmetric carbon peaks with GL(30) by setting the range of full width at half maximum to 1.2–2 eV. The FWHM of oxygenated carbon peaks, deconvoluted O1*s* peaks, and deconvoluted N1*s* peaks are set to 1.8-2.2 eV.[47] The *sp*$^3$-C peak is generally shifted by 0.7-1 eV from *sp*$^2$-C and hydroxyl/ether, carbonyl and carboxylic groups are shifted approximately 1.5, 3, and 4.5 eV higher, respectively.[47]

The Raman spectra of all samples were collected using a Renishaw *Invia* Raman spectrometer, UK. A 514 nm laser with a power of 0.04 mW, a 1800 grating spectrometer and a 50X objective lens were used to collect the spectra, with accumulation time of 5 s. The background correction of each spectrum is carried out using the WIRE software, provided by the instrument, using polynomial baseline fit with 5$^{th}$ order and normalized. The Raman spectra of grown materials were deconvoluted using MATLAB. The D-peak was fitted with a Lorentzian line-shape and the G-peak with a Breit-Weigner-Fano line-shape. The fitted spectra are supplied in the Supporting Information.

**Electrochemical measurements**. The electrochemical performances of the samples were carried out in 2-electrode configuration using Swagelok Cell (SKU: ANR-B01, Singapore). and 6M KOH used as an electrolyte. The hydrophobic PP membrane is modified by a two-step process: soaked with acetone at 20

°C for 5 min and followed by aqueous 6M KOH solution at 20 C, and used after 1 hr.[48] The cell was assembled by sandwiching separator-soaked-electrolytes between the carbon foam electrodes grown on carbon paper. For the electrochemical testing, carbon foams and modified separator were dipped into the electrolyte solution for 1hr. Cyclic voltammogram, charge-discharge test and electrochemical impedance spectra were recorded using a PALMSENS electrochemical workstation. Prior to the electrochemical test, cyclic voltammetry of assembled device was conducted within the electrochemical stable voltage at 100 mV/s for 1000 times. The cyclic voltammetry at different scan rates ranging from 20 to 1000 mV/s and charge-discharge at different current densities of 150 to 500 µA were carried out. The areal capacitance is calculated using the equation: $C_{areal} = \frac{\int I dV}{A.v.V}$, where $I$ is the current, $v$ is the scan rate, $A$ is the geometric area of the electrode and $V$ is the voltage of the device. The areal capacitance of the device from charge-discharge profile is estimated using the relation of $C_{areal} \frac{I_d . t_d}{(A.V)}$, where $I_d$ and $t_d$ are the discharge current and discharge time, respectively. Single electrode capacitance = 4 × device capacitance. The volumetric capacitance of electrode materials is estimated by dividing the areal capacitance by the total height of two carbon nanofoam electrodes. The electrochemical impedance spectroscopy is conducted in the frequency range of 1 Hz to 0.1 MHz at open circuit potential with a 10 mV *a.c.* perturbation. The energy density ($E_A$) of the device is calculated via the equation: $E_A = \frac{C_{dl} V^2}{2}$, where $C_{dl}$ is the double layer capacitance at 120 Hz obtained from the areal capacitance versus frequency plot. The relaxation time constant is calculated from the impedance spectra at 120 Hz using the equation: $\tau_{RC} = \frac{-Z'}{2\pi f Z''}$, where Z' and Z'' are the real and imaginary components of impedance.

**Supporting information.**

Additional SEM, XPS, Raman spectra, fitted curves with fitting details, and electrochemical results are provided in supporting information.

**Authors contributions**

S.G. planned and conceptualized the work, did the synthesis and characterizations of the materials, and wrote the manuscript. A.M., D.O. and A.M. assisted in PLD deposition. M.R. assisted in the electrochemical measurements and analysis. F.G. did the XPS measurements under the supervision of G.B. G.D. did the TEM, STEM and EELS measurements. V.R. helped in Raman spectroscopic analysis. D.D. and C.S.C. gave useful comments for the conceptualization of the work. All authors edited the manuscript and approved the final version of the manuscript.

**Notes**

The authors declare no competing financial interest.

**Acknowledgements**

S.G thank Horizon Europe (HORIZON) for the Marie Sklodowska-Curie Fellowship (grant no. 101067998-ENHANCER). Carlo S. Casari acknowledges partial funding from the European Research Council (ERC) under the European Union's Horizon 2020 Research and Innovation Program ERC Consolidator Grant (ERC CoG2016 EspLORE Grant Agreement 724610, website: www.esplore.polimi.it). The PhD scholarship of A.M. and part of this research study pertain to the Energy for Motion project of the Department of Energy of Politecnico di Milano, funded by the Italian Ministry of Education, University and Research (MIUR) through the Department of Excellence grant 2018–2022. Carlo S. Casari also acknowledges funding by the project funded under the National Recovery and Resilience Plan (NRRP), Mission 4 Component 2 Investment 1.3 Call for Tender 1561 of 11.10.2022 of Ministero dell'Università e della Ricerca (MUR), funded by the European Union NextGenerationEU Award Project Code PE0000021, Concession Decree 1561 of 11.10.2022 adopted by Ministero dell'Università e della Ricerca (MUR), CUP D43C22003090001, Project "Network 4 Energy Sustainable Transition (NEST)". A. M. and D. O. acknowledge the financial support of the European Union Next Generation EU and Italian Ministry of University and Research as part of the PRIN 2022 program (project ID:2022N5JBHT, CUP D53D23002840006). We thank Alessandro Zani, former PhD student at NanoLab, POLIMI for the MATLAB program for fitting the Raman spectra by BWF function. We also thank Alberto Calloni for the useful discussion on XPS analysis.

**Conflict of interests**

The authors declare no conflict of interest.

**Data Availability statement**

All the data of this study are available in the main manuscript and the Supplementary Information.

**References:**

[1] I. Kim, C. Shim, S. W. Kim, C. Lee, J. Kwon, K. Byun, U. Jeong, *Adv. Mater.* **2023**, *35*, 2204912.

[2] H. Mahajan, K. U. Mohanan, S. Cho, *Nano Lett.* **2022**, *22*, 8161.

[3] T. Lin, I.-W. Chen, F. Liu, C. Yang, H. Bi, F. Xu, F. Huang, *Science* **2015**, *350*, 1508.

[4] C. Casari, A. Bassi, In *Oxide Nanostructures*, Pan Stanford Publishing, **2014**, pp. 99–114.

[5] A. J. Haider, T. Alawsi, M. J. Haider, B. A. Taha, H. A. Marhoon, *Opt. Quantum Electron.* **2022**, *54*, 1.

[6] B. Bayatsarmadi, Y. Zheng, C. S. Casari, V. Russo, S.-Z. Qiao, *Chem. Commun.* **2016**, *52*, 11947.

[7] A. Maffini, A. Pazzaglia, D. Dellasega, V. Russo, M. Passoni, In *Nanoporous Carbons for Soft and Flexible Energy Devices*, **2022**, pp. 135–157.

[8] A. Hu, M. Rybachuk, Q.-B. Lu, W. W. Duley, *Appl. Phys. Lett.* **2007**, *91*, 131906.

[9] N. A. Kyeremateng, T. Brousse, D. Pech, *Nat. Nanotechnol.* **2017**, *12*, 7.

[10] S. Ghosh, J. Wang, G. Tontini, S. Barg, In *Flexible Supercapacitor Nanoarchitectonics*, **2021**, pp. 413–460.

[11]  S. Ghosh, S. R. Polaki, P. K. Ajikumar, N. G. Krishna, M. Kamruddin, *Indian J. Phys.* **2018**, *92*, 337.

[12]  S. Ghosh, S. K. Behera, A. Mishra, C. S. Casari, K. K. Ostrikov, *Energy & Fuels* **2023**, *37*, 17836.

[13]  Z. Supiyeva, X. Pan, Q. Abbas, *Curr. Opin. Electrochem.* **2023**, *39*, 101249.

[14]  A. Pazzaglia, A. Maffini, D. Dellasega, A. Lamperti, M. Passoni, *Mater. Charact.* **2019**, *153*, 92.

[15]  S. Ghosh, K. Ganesan, S. R. Polaki, T. R. Ravindran, N. G. Krishna, M. Kamruddin, A. K. Tyagi, *J. Raman Specroscopy* **2014**, *45*, 642.

[16]  A. Barinov, O. B. Malcioğlu, S. Fabris, T. Sun, L. Gregoratti, M. Dalmiglio, M. Kiskinova, *J. Phys. Chem. C* **2009**, *113*, 9009.

[17]  M. Pervolaraki, P. Komninou, J. Kioseoglou, A. Othonos, J. Giapintzakis, *Appl. Surf. Sci.* **2013**, *278*, 101.

[18]  E. Mohagheghpour, M. Rajabi, R. Gholamipour, M. M. Larijani, S. Sheibani, *Appl. Surf. Sci* **2016**, *360*, 52.

[19]  T. Danno, In *AIP Conference Proceedings*, AIP, **2004**, pp. 431–434.

[20]  A. P. Piedade, L. Cangueiro, *Nanomaterials* **2020**, *10*, 780.

[21]  S. I. Moseenkov, V. L. Kuznetsov, N. A. Zolotarev, B. A. Kolesov, I. P. Prosvirin, A. V. Ishchenko, A. V. Zavorin, *Materials* **2023**, *16*, 1112.

[22]  C. S. Casari, C. S. Giannuzzi, V. Russo, *Carbon* **2016**, *104*, 190.

[23]  A. Maffini, D. Orecchia, A. Pazzaglia, M. Zavelani-Rossi, M. Passoni, *Appl. Surf. Sci.* **2022**, *599*, 153859.

[24]  G. Sahoo, S. R. Polaki, S. Ghosh, N. G. Krishna, M. Kamruddin, K. (Ken) Ostrikov, *Energy Storage Mater.* **2018**, *14*, 297.

[25]  C. Casiraghi, A. C. Ferrari, J. Robertson, *Phys. Rev. B* **2005**, *72*, 085401.

[26]  A. C. Ferrari, S. E. Rodil, J. Robertson, *Phys. Rev. B* **2003**, *67*, 155306.

[27]  A. Maffini, A. Pazzaglia, D. Dellasega, V. Russo, M. Passoni, *Phys. Rev. Mater.* **2019**, *3*, 1.

[28]  S. Ghosh, S. Barg, S. M. Jeong, K. K. Ostrikov, *Adv. Energy Mater.* **2020**, *10*, 2001239.

[29]  D. Gandla, X. Wu, F. Zhang, C. Wu, D. Q. Tan, *ACS Omega* **2021**, *6*, 7615.

[30]  Y. Yoon, K. Lee, C. Baik, H. Yoo, M. Min, Y. Park, S. M. Lee, H. Lee, *Adv. Mater.* **2013**, *25*, 4437.

[31]  X. Liu, D. Lyu, C. Merlet, M. J. A. Leesmith, X. Hua, Z. Xu, C. P. Grey, A. C. Forse, *Science.* **2024**, *384*, 321.

[32]  J. Xue, Z. Gao, L. Xiao, T. Zuo, J. Gao, D. Li, L. Qu, *ACS Appl. Energy Mater.* **2020**, *3*, 5182.

[33]  N. Islam, J. Warzywoda, Z. Fan, *Nano-Micro Lett.* **2018**, *10*, 9.

[34]  Z. Zhang, Z. Wang, F. Wang, T. Qin, H. Zhu, P. Liu, G. Zhao, X. Wang, F. Kang, L. Wang, C. Yang, *Small* **2023**, *19*, 2300747.

[35]    J. R. Miller, R. A. Outlaw, B. C. Holloway, *Science* **2010**, *329*, 1637.

[36]    Y. Rangom, X. (Shirley) Tang, L. F. Nazar, *ACS Nano* **2015**, *9*, 7248.

[37]    Y. J. Kang, Y. Yoo, W. Kim, *ACS Appl. Mater. Interfaces* **2016**, *8*, 13909.

[38]    V. K. Mariappan, K. Krishnamoorthy, S. Manoharan, P. Pazhamalai, S. Kim, *Small* **2021**, *17*, 2102971.

[39]    Z. J. Han, C. Huang, S. S. Meysami, D. Piche, D. H. Seo, S. Pineda, A. T. Murdock, P. S. Bruce, P. S. Grant, N. Grobert, *Carbon N. Y.* **2018**, *126*, 305.

[40]    N. Frese, S. Taylor Mitchell, C. Neumann, A. Bowers, A. Gölzhäuser, K. Sattler, *Beilstein J. Nanotechnol.* **2016**, *7*, 2065.

[41]    Z. G. Neale, M. J. Lefler, J. W. Long, D. R. Rolison, M. B. Sassin, R. Carter, *Nanoscale* **2023**, *15*, 16924.

[42]    Z. Li, S. Gao, H. Mi, C. Lei, C. Ji, Z. Xie, C. Yu, J. Qiu, *Carbon N. Y.* **2019**, *149*, 273.

[43]    S. Nufer, P. Lynch, M. Cann, M. J. Large, J. P. Salvage, S. Víctor-Román, J. Hernández-Ferrer, A. M. Benito, W. K. Maser, A. Brunton, A. B. Dalton, *ACS Omega* **2018**, *3*, 15134.

[44]    N. Muchlisha, D. M. Widjonarko, T. E. Saraswati, *J. Phys. Conf. Ser.* **2023**, *2556*, 012006.

[45]    M. K. Hoque, J. A. Behan, J. Creel, J. G. Lunney, T. S. Perova, P. E. Colavita, *Front. Chem.* **2020**, *8*, 593932.

[46]    M. Ayiania, M. Smith, A. J. R. Hensley, L. Scudiero, J.-S. McEwen, M. Garcia-Perez, *Carbon N. Y.* **2020**, *162*, 528.

[47]    M. Smith, L. Scudiero, J. Espinal, J.-S. McEwen, M. Garcia-Perez, *Carbon* **2016**, *110*, 155.

[48]    A. Ciszewski, B. Rydzyńska, *J. Power Sources* **2007**, *166*, 526.

# SUPPORTING INFORMATION

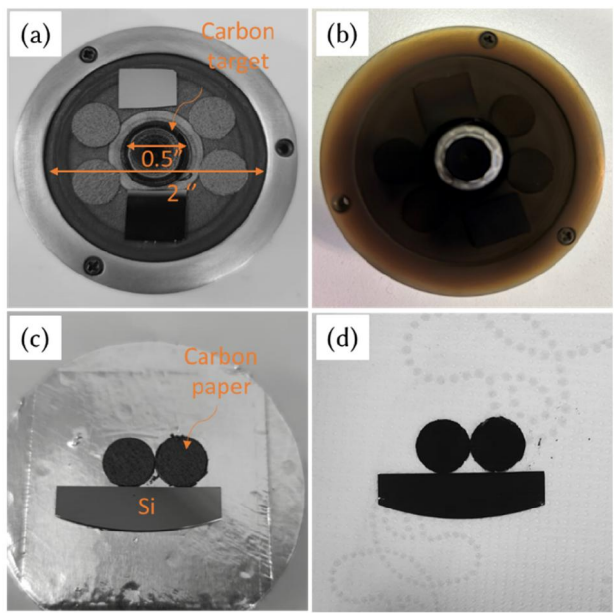

*Figure S1:* Photograph of target holder (a) before and (b) after carbon nanofoam deposition. Photograph of (c) substrate holder with the substrate before deposition and (d) substrate after deposition.

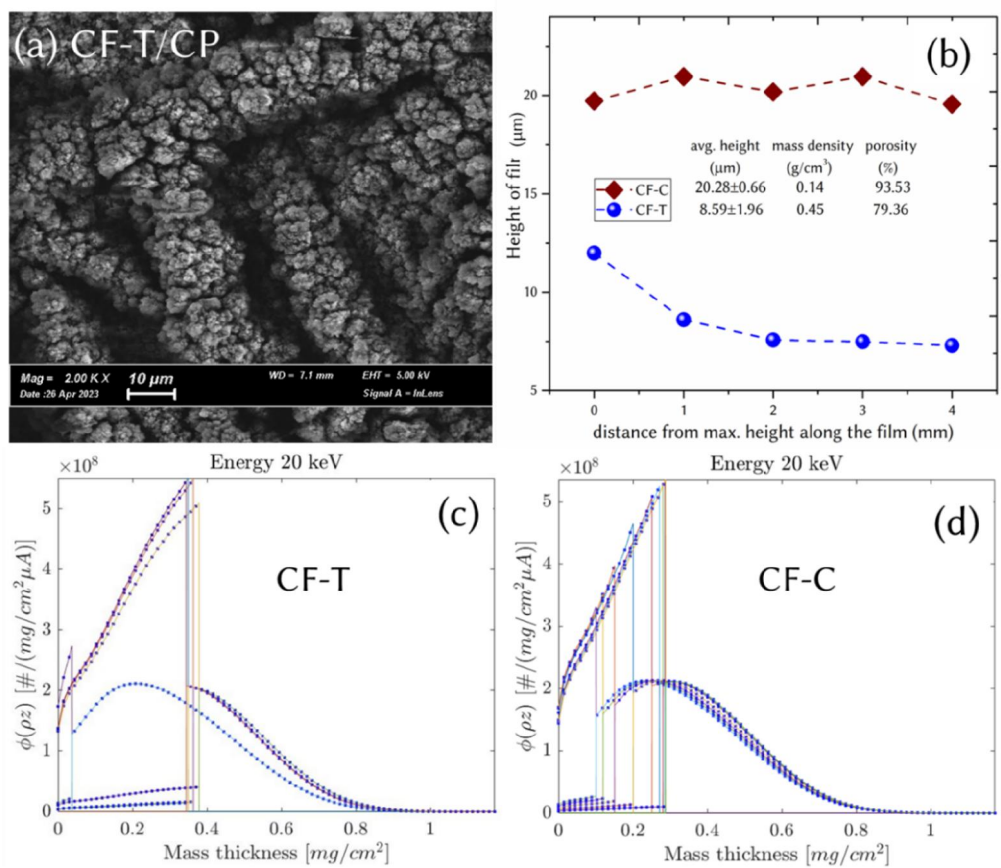

*Figure S2:* (a) scanning electron micrograph of CF-T grown on carbon paper, CP. (b) plot of height of the CF-T and CF-C film. Quantified average height, mass density and porosity of the films are provided in the inset of plot. (b) Simulated data of mass-density calculation of carbon nanofoam (c) CF-T and (d) CF-C using EDDIE software. The y-axis of plot (c-d) represent X-ray generation distribution function ($\phi(\rho z)$) in samples and each colored line corresponds to the iteration.

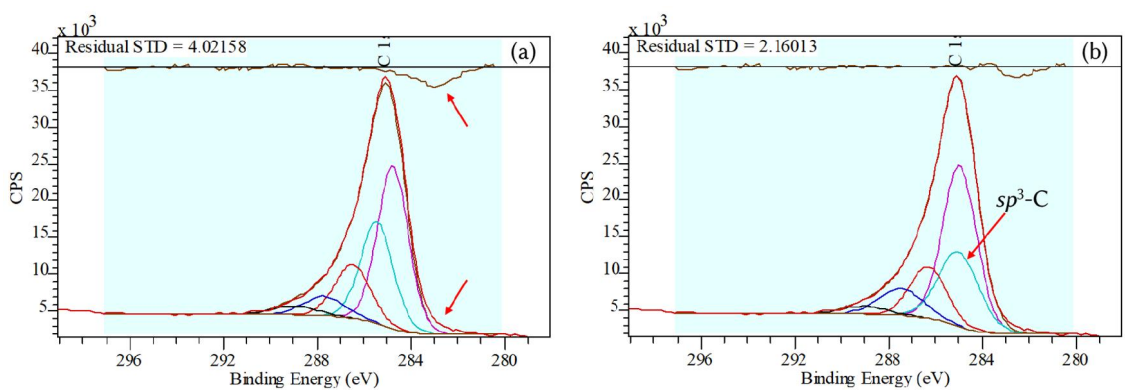

*Figure S3:* Casa XPS fitted C1*s* spectra without considering the peak *sp*-C at binding energy at around 283.8 eV. (a) Deconvoluted C1*s* spectra without *sp*-C peak shows higher residual STD of 4.02158. (b) fitted C1*s* spectra with deconvoluted peaks and without *sp*-C. In this fit, one peak ($sp^3$-C peak – sky blue color) lies under another peak ($sp^2$-C peak - pink colour). This fitting is not scientifically reliable. In the figure CPS stand for count per second and STD stand for standard deviation.

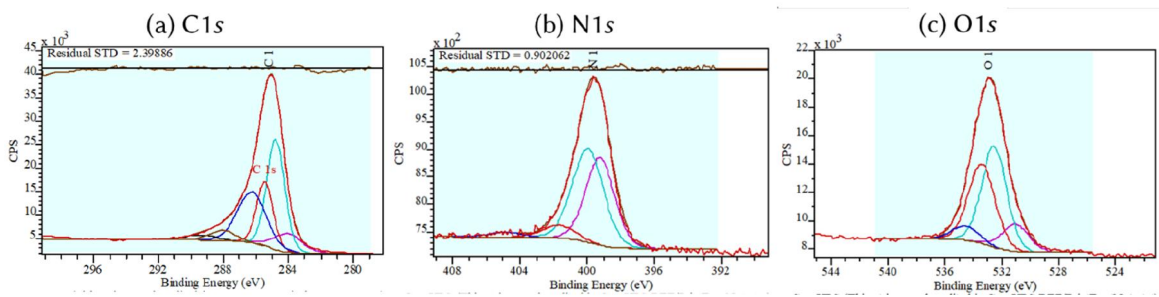

Figure S4. Casa XPS fitted High resolution (a) C1s, (b) N1s and (c) O1s spectra with deconvoluted peaks of CF-C.

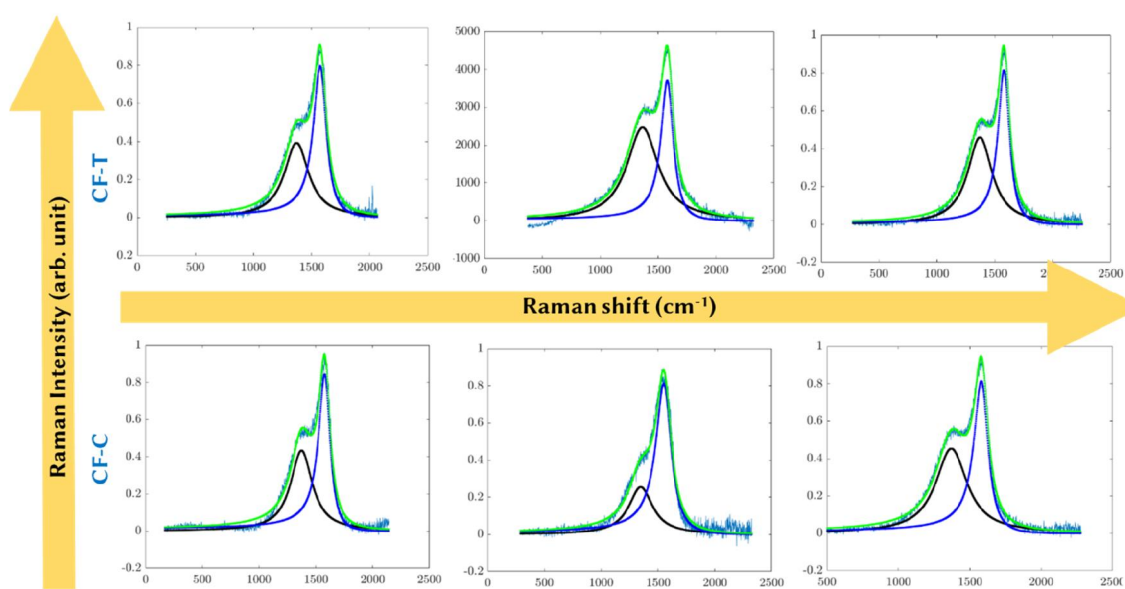

Figure S5: Fitted 1st-order Raman spectra of (top panel) CF-T and (bottom panel) CF-C taken at different position/samples using the MATLAB program. D-peak is Lorentzian shape and G-peak is fitted with BWF-lineshape

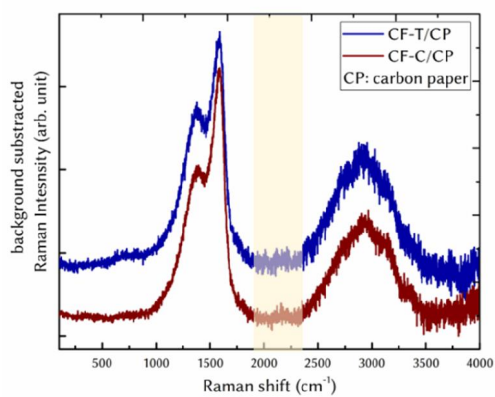

*Figure S6:* Raman spectra of CF-T and CF-C on carbon paper. A very weak *sp*-band is also observed for the sample on carbon paper, which is marked by highlighting region along with the other prominent peaks.

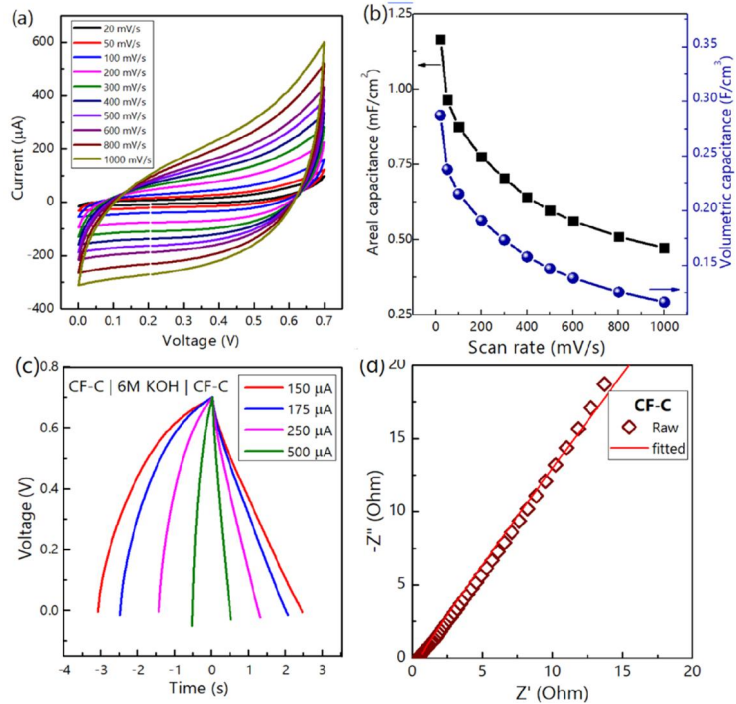

*Figure S7:* Electrochemical performance of CF-C symmetric device. (a) cyclic voltammogram at different scan rates, (b) areal and volumetric capacitance with respect to scan rate. (c) charge-discharge profile at different current. (d) Nyquist plot and fitted data with equivalent electric circuit.

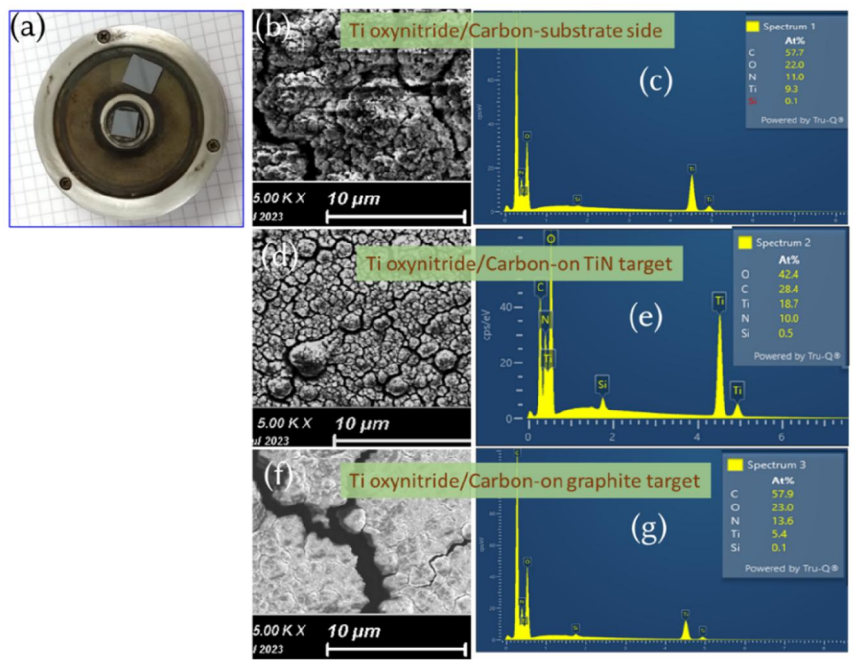

*Figure S8:* (a) Optical micrograph of the target side, where a Si substrate is placed on the graphite target and the TiN target. Scanning electron micrograph and corresponding EDX spectra of oxynitride/carbon composite prepared on the (b-c) substrate-side, (d-e) on TiN in target side and (f-g) on graphite in target side.

22